\documentclass[reprint,amsmath,amssymb,aps]{revtex4-2}
\usepackage{graphicx}

\usepackage{dcolumn}
\usepackage{bm}

\usepackage[dvipsnames]{xcolor}

\usepackage{amssymb}
\usepackage{amsmath}
\usepackage{epsfig}
\usepackage{chemarr}
\usepackage{url}
\usepackage{natbib}
\usepackage{color}

\usepackage[defaultcolor=red, final]{changes}

\begin{document}

\title{Physical networks become what they learn}

\author{Menachem Stern$^{1,2}$, Marcelo Guzman$^{1}$, Felipe Martins$^1$, Andrea J. Liu$^{1,3}$ and Vijay Balasubramanian$^{1,3,4}$}

\affiliation{$^1$Department of Physics and Astronomy, University of Pennsylvania, Philadelphia, PA 19104}

\affiliation{$^2$AMOLF, Science Park 104, 1098 XG Amsterdam, The Netherlands}

\affiliation{$^3$Santa Fe Institute, 1399 Hyde Park Road,
Santa Fe, NM 87501, USA}

\affiliation{$^4$Rudolf Peierls Centre for Theoretical Physics, University of Oxford, Oxford OX1 3PU, UK}

\date{\today}

\begin{abstract} 
Physical networks can develop tuned responses, or functions, by design, by evolution, or by learning via local rules. In all of these cases, tunable degrees of freedom characterizing internal interactions are modified to lower a \emph{cost} penalizing deviations from desired outputs. An important class of \added{such networks follows dynamics that minimize a global physical quantity}, or Lyapunov function with respect to physical degrees of freedom. \added{In such networks}, learning is a \emph{double optimization} process, in which two quantities -- one defined by the task and the other prescribed by physics --  are minimized with respect to different but coupled sets of variables. Here we show how this \added{learning} process couples the high-dimensional \emph{cost landscape} to the \emph{physical landscape}, linking the physical and cost Hessian matrices.  Physical responses of trained networks to random perturbations thus reveal the functions to which they were tuned. Our results, illustrated using electrical networks with adaptable resistors, are generic to \added{networks} that \added{perform} tasks \added{in the linear response regime}.
\end{abstract}

\maketitle


\added{Many physical networks have reciprocal edges, so that the interactions of nodes $i$ with $j$ and $j$ with $i$ are the same. Such networks can reach equilibrium or steady state by minimizing a global scalar function, or Lyapunov function, of the physical degrees of freedom.}
For example, mechanical networks minimize their (free) energy to achieve force balance on every node. Likewise, electrical resistor networks minimize dissipated power subject to current or voltage boundary conditions~\cite{vadlamani2020physics}, or, equivalently, satisfy Kirchhoff's laws. This minimization produces specific voltages at network nodes, and can be understood as a form of physical computation~\cite{landauer1987computation,piccinini2015physical, jaeger2023toward}. Such physical networks can develop desired linear responses to perturbations by simultaneously minimizing a cost function specifying the desired response. For example, electrical networks that use local rules that approximate gradient descent to learn desired computations~\cite{scellier2017equilibrium,kendall2020training,stern2021supervised,kaspar2021rise, stern2023learning, lopez2023self, anisetti2023learning, anisetti2024frequency} can be made in the laboratory~\cite{dillavou2022demonstration, wycoff2022learning, stern2022physical, dillavou2024machine, stern2024training}, and are arguably the simplest systems in which collective learning emerges from local dynamical processes.  

Systems \added{that optimize both a cost function and a physical Lyapunov function} have physical imprints of the \added{double optimization} process \added{in their linear response}: soft modes aligned with learned tasks~\cite{machta2013parameter, anisetti2023emergent,stern2024physical} appear in the dynamics. Here we elucidate the precise connection between the physical \added{linear response} of the network and its cost function.  

We show that in electrical networks that satisfy constraints on their responses to weak perturbations, the Hessian in the \emph{physical landscape}, characterizing curvatures around the minimized dissipated power in the space of node voltages, is directly related to the Hessian in the \emph{cost landscape}, characterizing curvatures around the minimized cost function in the space of adaptable degrees of freedom, or network parameters, namely the edge conductances. As in deep neural networks~\cite{sagun2016eigenvalues,sagun2017empirical, wei2019noise,sabanayagam2023unveiling}, the highest eigenmodes of the cost Hessian (directions of high curvature in the cost landscape) correspond to parameter changes that maximally impair performance of learned tasks. We derive an equation relating these stiff eigenmodes, after a transform determined by the network topology, to soft modes of the physical Hessian (directions of low curvature in the physical landscape).

Calculating the cost Hessian requires knowing the task. By contrast, the physical Hessian is a network property that can be measured \textit{via} responses to random perturbations~\cite{chen2010low}. Our results imply that the physical Hessian gives insight into an adaptable network's function \emph{without knowing the task}. This provides a new tool for understanding adaptable systems.  While our exposition is focused on electrical resistor networks, our results are general to all physical networks that achieve desired linear responses via a \added{double optimization} process.

\added{For networks that learn tasks via a contrastive learning procedure using local rules (such as electrical contrastive local learning networks that have been realized in the lab~\cite{dillavou2022demonstration,dillavou2024machine}), we derive a relation between the cost and physical Hessians during the learning process.}

\emph{Physical and cost Hessians} -- Consider a linear electrical resistor network, with $N$ nodes indexed by $\mathbf{a}$, carrying voltages $\mathbf{V_a}$ collected into a voltage vector $\mathbf{V}$. Nodes are connected by $N_e$ edges indexed by 
$i$, of conductance $\kappa_i$. We use bold font for dynamical physical degrees of freedom (physical DOF; node voltages), and Greek font for adaptive degrees of freedom (adaptive DOF; edge conductances). Electrical resistor networks minimize total power dissipation. Because our networks are linear, the dissipated power is 
\begin{equation}
\begin{aligned}
\mathbf{P}= \frac{1}{2}\mathbf{V}^{T} \mathbf{H} \mathbf{V}
\end{aligned}
 \label{eq:Power},
\end{equation}
with a symmetric \emph{physical} $N\times N$ Hessian matrix, 
\begin{equation}
 \mathbf{H_{ab}}=
 \frac{\partial^2 \mathbf{P}}{\partial\mathbf{V_a} \partial\mathbf{V_b}} 
 \equiv
 2 \left[ \mathbf{\Delta}^{T} \kappa  \mathbf{\Delta}\right]_{\mathbf{ab}},
 \label{eq:physHessian}
\end{equation}
where $\mathbf{a,b}$ index  $N$ nodes; $i$, $j$ index $N_e$ edges \added{connecting pairs of nodes}; $\kappa$ is a diagonal matrix of edge conductances with $\kappa_{ij} = \kappa_i \, \delta_{ij}$; and  $[ \cdots ]_{\mathbf{ab}}$ indicates a component of the matrix in the parentheses. $H$ is also the graph Laplacian, but we refer to it generically as the Hessian. $\mathbf{\Delta}$ is an \textit{incidence matrix} specifying the network geometry, with $\mathbf{\Delta}_{j \mathbf{a}} = 1,0,-1$ depending on whether edge $j$ is incoming, disconnected, or outgoing to node $\mathbf{a}$, with arbitrarily assigned edge directions. Thus, the physical Hessian $\mathbf{H}$ depends on the adaptive conductances $\kappa_i$.

When external currents $\mathbf{I_a}$, collected into a vector $\mathbf{I}$, are applied to network nodes,  the network minimizes the \emph{free state} power dissipation~\cite{vadlamani2020physics}, i.e., total power subject to inputs: $\mathbf{P}^F= \mathbf{V}^{T} (\frac{1}{2}\mathbf{H} \mathbf{V} - \mathbf{I})$. The minimum of the free state power is achieved by the free state voltage values:
\begin{equation}
\begin{aligned}
\mathbf{V}^{F}(\kappa;\mathbf{I})=\mathbf{H}^{-1}(\kappa) \, \mathbf{I}.
\end{aligned}
 \label{eq:PF}
\end{equation} 
For electrical networks, the Hessian in (\ref{eq:physHessian}) is not strictly invertible because it has a zero mode -- shifting all voltages by a constant leaves dissipated power unchanged. In practice we include an additional ground node with constrained voltage $V=0$ (see SI Note 1); this removes the zero mode and renders the Hessian invertible.

In this context, learning a target response amounts to adapting conductances to satisfy a constraint on the free state response $c(\mathbf{V}^F)=0$~\cite{mehta2019high}.
We quantify this objective by a scalar cost function $C$, the square of the constraint:
\begin{equation}
C\equiv \frac{1}{2} c^2 \, .
\label{eq:costdef}
\end{equation}
Multiple learning constraints \added{(distinct input-output pairs in the machine learning literature)} specified by $c_i$ are codified by a joint cost $C = \frac{1}{2}\sum_i c_i^2$. For simplicity \added{we consider tasks with a single constraint} and show in SI Note 2 that our results \added{on general relations between the physical and cost Hessian extend} to multiple \added{constraints}, \added{exemplified in a linear regression task.} The process of modifying the adaptable conductances to minimize the cost function $C$ is called supervised learning when training examples with labels are provided. The cost measures how well the network reproduces labels, or, more generally, the desired response.

Learning terminates when the system reaches a minimum of the cost function, so that both the cost function and its gradient with respect to $\kappa_i$ vanish.  The cost landscape is then locally described to lowest nonvanishing order by the \emph{cost Hessian}, an $N_e \times N_e$ matrix
\begin{equation}
\mathcal{H}_{ij}\equiv \frac{d^2 C}{d \kappa_i d \kappa_j}.
\label{eq:costHessian}
\end{equation}
where $N_e$ counts adaptable parameters. The cost Hessian is studied in machine learning approaches~\cite{dauphin2014Indentifying}. For a network well-trained for $n_T$ \added{independent constraints}, the cost Hessian has $n_T$ finite eigenvalues~\cite{sagun2016eigenvalues, sagun2017empirical, wei2019noise}.


\emph{Relation between the physical and cost Hessians}-- 
The physical and cost Hessians have different dimensions and units, but we will see that they are nevertheless related. We can rewrite the cost Hessian as
\begin{equation}
\begin{aligned}
\mathcal{H}_{ij}=\frac{d c}{d \kappa_i} \frac{d c}{d \kappa_j} + c \frac{d^2 c}{d\kappa_i d\kappa_j} \equiv\mathcal{H}_{ij}^{\mathrm{sat}} + \mathcal{H}_{ij}^{\mathrm{dyn}} \, ,
\end{aligned}
 \label{eq:CHess0}
\end{equation}
a sum of
the \emph{satisfied} term $\mathcal{H}_{ij}^{\mathrm{sat}} \equiv \frac{d c}{d \kappa_i} \frac{d c}{d \kappa_j}$ and the \emph{dynamic} term $\mathcal{H}_{ij}^{\mathrm{dyn}} \equiv c \frac{d^2 c}{d\kappa_i d\kappa_j}$. $\mathcal{H}_{ij}^{\mathrm{dyn}}$ vanishes if the network achieves learning since $c=0$.  

From (\ref{eq:CHess0}) we see that for one constraint, $\mathcal{H}_{ij}^{\mathrm{sat}}$ is rank-1, as it is an outer product of a vector $g_i \equiv \frac{d c}{d \kappa_i}$ with itself. $g_i$ is proportional to an eigenmode of $\mathcal{H}_{ij}^{\mathrm{sat}}$.  

To relate the physical and cost Hessians,  recall that the constraint $c(\mathbf{V}^F)$ is an explicit 
function of the free state response, so that
\begin{equation}
\begin{aligned}
g_i= 
\frac{\partial c}{\partial \mathbf{V}^F} \cdot 
\frac{d\mathbf{V}^{F}}{d\kappa_i} 
= 
-\frac{\partial c}{\partial \mathbf{V}^F}  \cdot 
\mathbf{H}^{-1}\frac{d\mathbf{H}}{d\kappa_i} \mathbf{H}^{-1} \mathbf{I}
\end{aligned}
 \label{eq:FSD1}.
\end{equation}
where we used (\ref{eq:PF}) along with the formula for the derivative of the inverse of a matrix. Finally, applying the  $\kappa_i$ derivative to the physical Hessian in Eq.~\ref{eq:physHessian} we arrive at 
\begin{equation}
g_i= - 2 (\frac{\partial c}{\partial \mathbf{V}^F} \cdot \mathbf{H}^{-1} \mathbf{\Delta}_i^T) (\mathbf{\Delta}_i \mathbf{H}^{-1} \mathbf{I})  \, ,
\label{eq:gi}
\end{equation}
where by ${\bf \Delta}_i$ we mean the the ith row of ${\bf \Delta}$.  So ${\bf \Delta}_i$ is a vector of length $N$ whose entries are 1 for the node that edge $i$ enters, $-1$ for the node it leaves, and $0$ for the other nodes.  We find that
\begin{equation}
\added{\mathcal{H}_{ij}^{\mathrm{sat}} = g_i g_j ,}
 \label{eq:FSD2}
\end{equation}
\added{where $\mathcal{H}_{ij}^{\mathrm{sat}}$ contains four inverse powers of $\mathbf{H}$.} The additional factors of $\mathbf{\Delta}$ from (\ref{eq:gi}) implement a transform determined by the network topology,  transporting the inverse physical Hessian into network parameter space.

Eq.~\ref{eq:gi} and~\ref{eq:FSD2} encode our main result, a remarkable connection between the physical and cost landscapes when the constraint is satisfied thus minimizing cost: the curvature around the solution in the \emph{cost} landscape is directly related to the curvature around the minimum of the \emph{physical} landscape.  This connection arises in electrical networks because power must be minimized to satisfy Kirchhoff's law at every node while cost is decreased; this couples the two landscapes in their respective spaces. 

The satisfied cost Hessian in (\ref{eq:FSD2}) is an outer product of  $g_i$ (the gradient constraint $c$ wrt $\kappa_i$) with itself. Thus, for one constraint, we can think of a \emph{reduced} cost Hessian of rank $1$, with one non-zero eigenvalue and corresponding eigenmode $g_i$ associated with the learned task.
Moreover, from the definition of the cost function (\ref{eq:costdef}),
\begin{equation}
  \frac{d C}{d \kappa_i}=cg_i
\end{equation}

As the cost vanishes,  its gradient is proportional to the nontrivial cost Hessian eigenmode.

Our analysis generalizes to physical systems that minimizes a scalar physical global quantity, including mechanical networks that minimize energy or free energy (SI Note 4), or not only linear but also nonlinear flow and electrical networks that minimize dissipated power or, in the nonlinear case, a ``co-content"~\cite{parodi2018linear, kendall2020training}. The reason is that, for weak inputs, the dynamics are controlled by a quadratic energy functional with a fixed Hessian, and hence linear responses, leading to equations of the same form as those we have solved. 

To relate the physical and cost Hessian eigenmodes, we rotate to natural coordinates of the physical Hessian:

\begin{equation}
\begin{aligned}
H = \bm v^T \Lambda \bm v \quad;\quad \mathbf{i} \equiv \bm v \mathbf{I} \quad;\quad \mathbf{o} \equiv \bm v \frac{dc}{d\mathbf{V}^F}.
\end{aligned}
 \label{eq:Hrot}
\end{equation}

Here the rows of matrix $\bm v$ are  eigenmodes of the physical Hessian, $\Lambda$ is a diagonal matrix of  associated  eigenvalues $\mathbf{\lambda}>0$, and $\mathbf{i},\mathbf{o}$ are the input current and  constraint gradient rotated to the physical Hessian reference frame. 

In this coordinate system

\begin{equation}
\begin{aligned}
g_i= - 2 \sum_{\mathbf{\mu\nu}} \Big(\frac{\mathbf i}{\lambda}\Big)_{\mathbf{\mu}} \Big(\frac{\mathbf o}{\lambda}\Big)_{\mathbf{\nu}} (\mathbf{\Delta}_i \bm v_{\mathbf{\mu}}^T) (\mathbf{\Delta}_i \bm v_{\mathbf{\nu}}^T) 
\end{aligned}
 \label{eq:Srot}.
\end{equation}
The vector $(\frac{\mathbf{i}}{\lambda})_{\mathbf{\mu}}$ is the input current projected onto the direction of eigenmode $\bm{v}_{\mathbf{\mu}}$, divided by the associated eigenvalue $\lambda_\mu$; similarly for $(\frac{\mathbf{o}}{\lambda})_{\mathbf{\mu}}$. $\mathbf{\Delta}_i {\bm v}^T_{\mathbf{\mu}}$ is the voltage difference across the $i$th edge for the $\mathbf{\mu}$th eigenvector.

\added{Intuitively, a relation arises between the cost and physical Hessians eigenmodes in trained systems because the highest eigenmodes of the cost Hessian {\it define} a desired {\it physical response}. This physical linear response $\mathbf{V}^F=\mathbf{H}^{-1}\mathbf{I}=\sum_\mu(\frac{\mathbf{i}}{\lambda})_{\mathbf{\mu}}\bm{v}^T_\mu$ is controlled by the inverse physical Hessian and dominated by its lowest eigenmodes. Thus, the lowest physical modes that project onto the desired response, having large $(\frac{\mathbf{o}}{\lambda})_{\mathbf{\mu}}$, are related to the highest eigenmodes of the cost Hessian.} 



We previously showed that \added{in systems learning linear responses}, the physical Hessian develops soft modes with a small eigenvalues \added{and large projections onto the desired responses~\cite{stern2024physical}}. \added{Note that soft modes that are not learned could exist if a network has certain symmetries or structural features. However, such unlearned modes are independent of the adaptive degrees of freedom, and generally do not project much onto the desired system response, and thus do not couple to $g_i$ or the cost Hessian.} \added{In simple cases where only one output constraint is learned, one low eigenmode arises with large projection on the constraint.} If its eigenvalue is much smaller than the rest, we can approximate (\ref{eq:Srot}) by one term,  as  discussed above for the cost Hessian of networks trained to achieve one constraint 
\begin{equation}
\begin{aligned}
g_i^{\mathrm{Approx}} = -2 \frac{\mathbf i_1 \mathbf o_1}{\lambda_1^2} (\mathbf{\Delta}_i \bm v_1^T)^2
\end{aligned}
 \label{eq:S2}.
\end{equation}
In this case we find that the stiff mode of the cost Hessian $\mathcal{H}$ has components proportional to the squared differences of node voltages belonging to the soft mode of the physical Hessian $\bf{H}$.  Note that $\bf{H}$ can be estimated from physical responses alone~\cite{chen2010low}, with no knowledge about the constraints the system was required to satisfy. This implies that without knowing $\mathbf{i},\mathbf{o}$ we can obtain a good approximation of the stiff mode of $\mathcal{H}^{\mathrm{sat}}$.  One does not need to specify the task -- to know the constraint or cost function--in order to learn which edges most strongly affect the error in a trained network. \added{It is more complicated to relate individual physical eigenmodes to constraints when multiple soft modes of $\bf{H}$ are relevant.  We leave these cases to future work.} 

\begin{figure}
\includegraphics[width=0.85\linewidth]{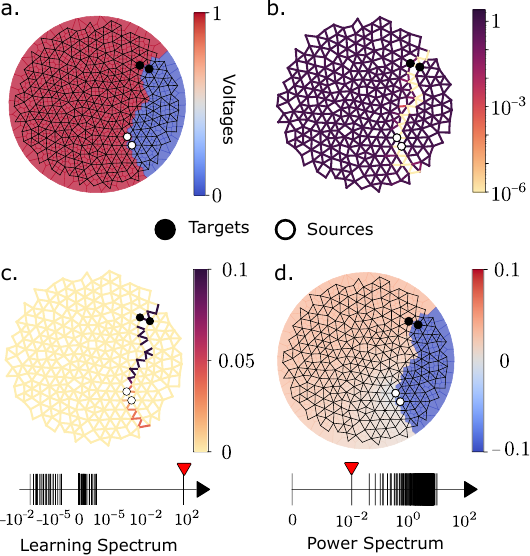}
\caption{Linear electric resistor networks have correlated modes in the physical and cost spaces. Here, the task requires the voltage drop between two output nodes (black) to equal to the voltage drop between two input nodes (white). (a) The  response of the network. Color interpolates between  node voltages. The response is partitioned into two sectors, with high uniform voltage and low uniform voltage, respectively. (b) The edge conductances; the network develops a crack of low conductance. (c) The stiff mode of the cost Hessian picks out the edges in the crack. (d) The soft mode of the physical Hessian reflects the voltage response and complements the stiff mode of the learning cost Hessian.
\label{fig:fig1}}
\end{figure}

Fig.~\ref{fig:fig1} shows a network trained so that a voltage drop applied between the two white input nodes yields an equal voltage drop between the two black output nodes. Here, inputs are applied voltages, rather than applied currents as discussed above. \added{In our analysis, inputs could be set as currents or voltages (relative to ground) {\it supplied to nodes}, or as fixed voltages \emph{across  edges}.  Also, we consider output constraints that do not explicitly depend on the adaptive conductances $\kappa_i$ (SI Note 1 and 5).  If the output constraints depend explicitly on $\kappa_i$ (e.g., for  power constraints),  or if they do not have a leading linear term in the response $\mathbf{V}^F$, our results are modified as discussed in SI Note 3.} In Fig.~\ref{fig:fig1}, training partitions the network into high voltage and low voltage sectors (Fig.~\ref{fig:fig1}a), separated by a crack of low conductance (Fig.~\ref{fig:fig1}b).

Fig.~\ref{fig:fig1}c shows that the stiff mode of the cost Hessian is spatially localized to the low-conductance crack; if these conductances are varied, the cost increases strongly. Likewise Fig.~\ref{fig:fig1}d shows that the soft mode of the physical Hessian reflects the voltage response in  Fig.~\ref{fig:fig1}a and complements the stiff mode of the cost Hessian. This is the soft response mode reached by learning.  Note that many other solutions, each with its own physical Hessian, are also possible~\cite{rocks2019limits} but following \cite{rocks2017designing,rocks2021hidden,stern2024physical}, systems trained for such linear responses typically find solutions with stiff cost eigenmodes and soft physical eigenmodes.
Significantly, the lowest non-trivial eigenmode of the physical Hessian (Fig.~\ref{fig:fig1}d), captures the network  response to the input, without knowledge of the expected input or output. Applying the  difference operator $\mathbf{\Delta}_i$ to this soft mode, we find that $(\Delta_i v_1)^2$ is  spatially localized to the crack; the soft mode reflects key edges associated with the stiff cost Hessian mode. The latter demarcates the boundary between sectors separated by the physical Hessian, which can also be revealed through persistent homology analysis of the trained network~\cite{rocks2020revealing,rocks2021hidden}.

Once the constraint is satisfied, we can approximate the cost Hessian as (\ref{eq:FSD2}):
\begin{equation}
\begin{aligned}
\mathcal{H}_{ij}^{\mathrm{Approx}} \approx 4
\frac{(\mathbf i_1 \mathbf o_1)^2}{\lambda_1^4} (\mathbf{\Delta}_i \bm v^T_1)^2 (\mathbf{\Delta}_j \bm v^T_1)^2
\end{aligned}
 \label{eq:CostHessTrained3}.
\end{equation}

This has rank 1 as the outer product of a vector. The trace equals the non-zero eigenvalue: $\lambda^\mathcal{L}_1 = {\rm Tr} (\mathcal{H}^{\mathrm{Approx}})$, relating the stiff cost  to the soft physical eigenvalue:
\begin{equation}
\begin{aligned}
\lambda_1^{\mathcal{L}} = 4 \sum_i g_i^2 \approx \frac{(\mathbf{i}_1 \mathbf{o}_1)^2}{\lambda_1^4} \sum_i(\mathbf{\Delta}_i \bm{v}^T_1)^4 \propto \lambda_1^{-4}
\end{aligned}
 \label{eq:CostHessTrained4}.
\end{equation}
These results are illustrated in SI Note 6 and Fig.~S3 with resistor networks trained on tasks similar to Fig.~\ref{fig:fig1}.

\begin{figure}
\includegraphics[width=0.95\linewidth]{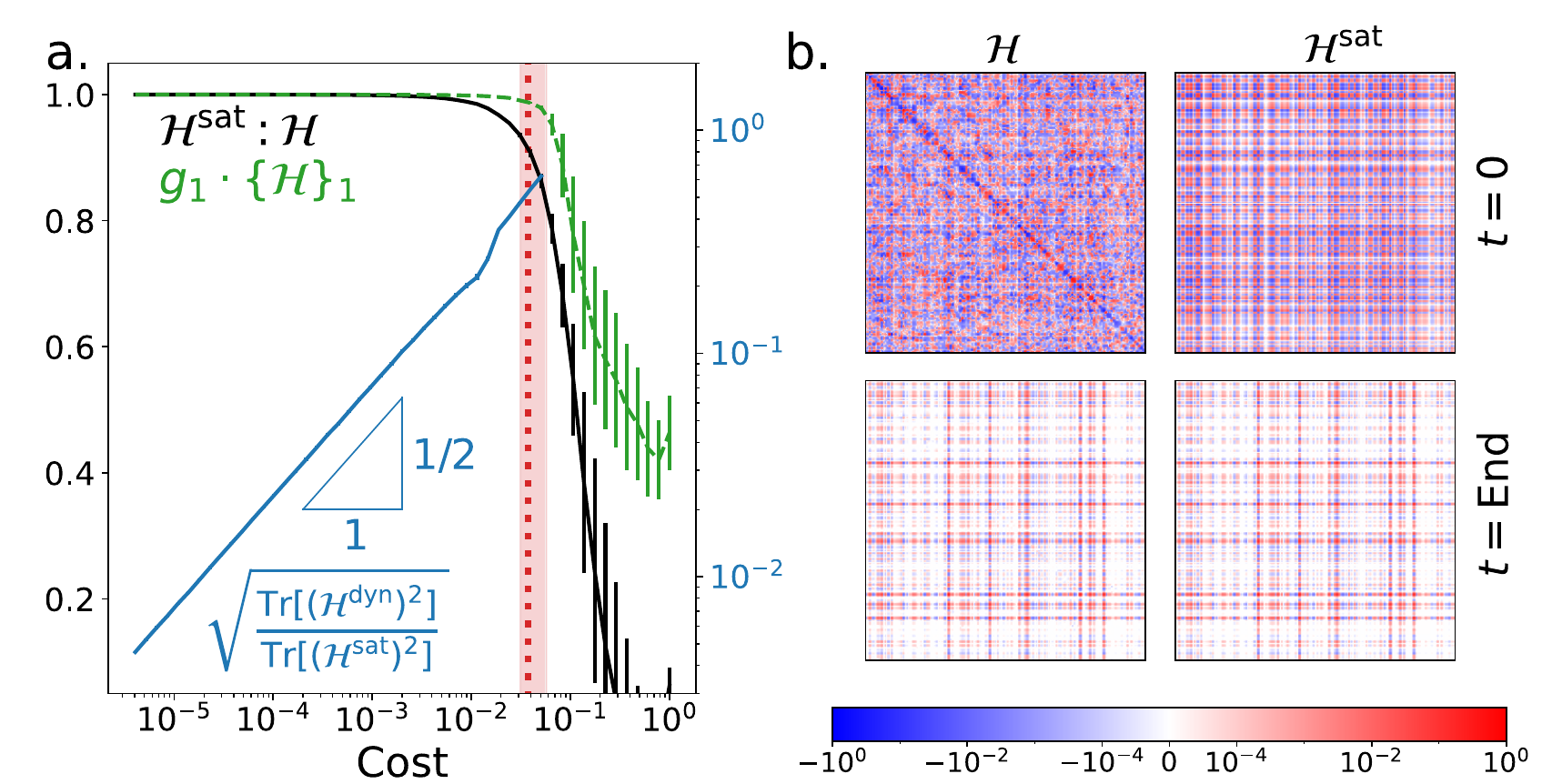}
\caption{Cost Hessian as cost $C$ decreases during the learning for a task constraining voltage drops between output nodes in response to currents applied at input nodes. (a) The cost Hessian is captured well by $\mathcal{H}^\mathrm{sat}$ (black curve), when $C$ is less than the scale set by the lowest physical eigenvalue $C \lesssim \lambda_1^{-2}$. Vertical red line =  median of $\lambda_1^{-2}$. Middle 80\% percentile of the range = red shaded region. Similarly, Below that same scale, the $g_i$ vector in Eq.~\ref{eq:gi} (green curve) captures the stiffest vector of the cost Hessian ($\{{\cal H}\}_1$). The ratio of dynamic to satisfied cost Hessians (blue curve) scales as $\sqrt{C}$, as predicted. (b) Comparison of the full and satisfied cost Hessians before and after the \added{learning} process. 
\label{fig:fig3}}
\end{figure}

\emph{Dynamical Hessian relations} -- We now examine how the two Hessians are related during learning, not just when the constraints have been satisfied. From  (\ref{eq:CHess0}):
\begin{equation}
\begin{aligned}
\mathcal{H}_{ij}&= g_i g_j + 4c L_i M_{ij} R_j + 4c L_j M_{ji} R_i \\
&+  4c  R_i R_j 
    \sum_{\mathbf{ab}} ({\bf \Delta}_i {\bf H}^{-1})_{\mathbf{a}}
        \frac{\partial^2 c}{\partial {\bf V}^F_{\mathbf{a}}  \partial {\bf V}^F_{\mathbf{b}}}
        ({\bf H}^{-1} {\bf \Delta}^T_j)_{\mathbf{b}}  
\end{aligned}
 \label{eq:CHess1}
\end{equation}
where we defined $ L_i \equiv \frac{\partial c}{\partial{\bf v}^F} \cdot {\bf H}^{-1} {\bf \Delta}_i^T$, $R_i \equiv {\bf \Delta}_i {\bf H}^{-1} {\bf I}$,
and $M_{ij} \equiv {\bf \Delta}_i {\bf H}^{-1} {\bf \Delta}_j^T$ (details in SI Note 7).  The last three terms arise from $\mathcal{H}_{ij}^{\mathrm{dyn}}$.  Recall that $\mathcal{H}_{ij}^{\mathrm{sat}}\sim \mathbf{H}^{-4}$.  The first two terms in  $\mathcal{H}_{ij}^{\mathrm{Dyn}}$ scale as  $c\mathbf{H}^{-3}\sim \sqrt{C}\mathbf{H}^{-3}$, while the last term scales as $c\mathbf{H}^{-4}\sim \sqrt{C}\mathbf{H}^{-4}$.  For systems trained with linear constraints, the last term vanishes because $\frac{\partial^2 c}{\partial {\bf V}^F_a  \partial {\bf V}^F_b} = 0$. Then
$\mathcal{H}^\mathrm{sat}$ dominates, $\mathrm{Tr}[(\mathcal{H}^{\mathrm{sat}})^2] \gg \mathrm{Tr}[(\mathcal{H}^{\mathrm{dyn}})^2]$ once the error is less than a scale set by the lowest physical Hessian eigenvalue, $C \ll \lambda_1^{-2}$.  For nonlinear constraints, $\mathcal{H}^{\mathrm{sat}}$ still dominates, at least for small enough error, $C \ll 1$.   The change of the physical Hessian $\bf{H}$ during learning is proportional to the cost function gradient. Thus, the physical Hessian $\mathbf{H}$ becomes nearly constant when  $C$ is low enough.

Fig.~\ref{fig:fig3} shows results for 300 networks trained with a linear constraint to produce specified voltage drops to applied input currents, and calculated $\mathcal{H}^\mathrm{sat}$ and $\mathcal{H}^\mathrm{dyn}$. The cost Hessian $\mathcal{H}$ is well described by $\mathcal{H}^\mathrm{sat}$ (Fig.~\ref{fig:fig3}a) -- the normalized double dot product $\mathcal{H}^\mathrm{sat} :\mathcal{H} \equiv \frac{\mathrm{Tr}[\mathcal{H}^\mathrm{sat} \mathcal{H}]}{\sqrt{\mathrm{Tr} [(\mathcal{H}^\mathrm{sat})^2]}\sqrt{\mathrm{Tr}[\mathcal{H}^2]}}$
approaches $1$ (black) when the cost is below a threshold $\lambda_1^{-2}$ set by the lowest physical Hessian eigenvalue, $\lambda_1$. Similarly, the physical approximation of $g_i$  (\ref{eq:gi}) approaches the stiffest cost eigenvector $\{{\cal H}\}_1$, generating its rank-1 reduced form.
Fig.~\ref{fig:fig3}a also shows that the ratio of the dynamic and satisfied terms scales with $\sqrt{C}$ (blue), as predicted. Fig.~\ref{fig:fig3}b  shows the cost Hessian for a network before and after training. Initially $\mathcal{H}^\mathrm{sat}$ approximates the cost Hessian poorly, but afterwards captures it well, with the same sparse, low-rank features.


\emph{Discussion} -- We derived relations between the physical and cost Hessians for networks responding to weak forces near equilibrium as they adapt to perform tasks. For simplicity, in the main text we focused on networks satisfying a single constraint, and thus have a rank-1 reduced cost Hessian. \added{In SI Note 2 we show how networks that satisfy $n_T$ constraints have reduced cost Hessians with rank up to $n_T$.}  
We also showed how, after learning, the high eigenmodes of the cost Hessian are related to low eigenmodes of the physical Hessian, and used examples to illustrate how these high modes are supported primarily on network edges that adjust to satisfy the constraints.  For a physical system that satisfies a single constraint, our results show that we should expect a single low physical eigenmode, separated by a gap from the higher modes, and directly related to a single high cost Hessian eigenmode. Then, the satisfied constraint can be discovered by examining the physical responses of the network around equilibrium without any prior knowledge of the task. Even when the low physical eigenmodes are not separated by a gap, the connection between the Hessians suggests strategies for learning about stiff cost Hessian eigenmodes from physical information only~\cite{Guzman2024Imprints}. While we treated the case of linear electric resistor networks, our results hold generally for steady-state physical networks that minimize a scalar cost specifying a linear response (SI Note 4). In other words, we find that, in a sense, \textit{physical networks become what they learn}.

Our approach applies to adaptable mechanical systems~\cite{pashine2019directed,stern2020supervised, arinze2023learning, altman2024experimental, mei2023memory}, including proteins, where correlations between physical structure and function \cite{halabi2009protein,tecsileanu2015protein,raman2016origins,mitchell2016strain,tlusty2017physical,rouviere2023emergence} appear in regions highly conserved over evolutionary times, suggesting their functional importance~\cite{tecsileanu2015protein,salinas2018coevolution}. Conserved regions in proteins were associated with slow collective modes--corresponding to low eigenmodes of the physical Hessian~\cite{husain2020physical}. Because our approach links the physical and cost Hessians for each system, it can give insight into individual systems as well as for an ensemble of solutions, e.g. a protein family, shedding light on idiosyncratic features that play functional roles~\cite{falk2023learning}. 

More broadly, circuits in the brain display structures conserved between individuals and  species~\cite{olivares2001species}. In many cases these conserved structures are evolved or learned adaptations to the environment or to tasks, for example, in circuits supporting vision \cite{ratliff2010retina,garrigan2010design,hermundstad2014variance}, audition \cite{lewicki2002efficient,smith2006efficient}, olfaction \cite{tecsileanu2019adaptation,krishnamurthy2022disorder}, and spatial cognition \cite{wei2015principle}. However, individual differences are also functionally important, e.g., for encoding  memories~\cite{vogel2004neural, barrett2004individual}.  Our findings linking structure and function in individual networks may help to explain how conserved network adaptations can be accompanied by substantial individual variations~\cite{rouw2010neural, shinkareva2012exploring}. Finally, the connections we have uncovered between the cost Hessian and physically measurable quantities suggest that we can apply the rich machine learning literature exploiting the cost Hessian to develop more efficient physical learning rules~\cite{dauphin2014Indentifying}.


We thank Eric Rouviere, Pratik Chaudhari, Benjamin Machta, Pankaj Mehta, and Arvind Murugan for insightful discussions. This work was supported by DOE Basic Energy Sciences through grant DE-SC0020963 (MS,MG), the UPenn NSF NRT DGE-2152205 (FM) and the Simons Foundation through Investigator grant \#327939 to AJL.  VB and MS were also supported by NIH CRCNS grant 1R01MH125544-01 and NSF grant CISE 2212519. M.G. acknowledges support from postdoctoral fellowships at the Center of Soft and Living Matter and the Data-Driven Discovery Initiative, both at the University of Pennsylvania. VB was supported in part by the Eastman Professorship at Balliol College, University of Oxford. This work was completed at the Aspen Center for Physics (NSF grant PHY-2210452).


\end{document}